\documentclass{mem}
\usepackage{natbib}\usepackage{txfonts}\usepackage{balance}
\usepackage{graphicx}
\usepackage[a4paper,breaklinks,dvipdfm]{hyperref}
\idline{75}{282}
\begin{document}
\def\teff{$T\rm_{eff }$}
\def\kms{$\mathrm {km s}^{-1}$}

\title{
BSS populations from the WFPC2 UV Survey
}

   \subtitle{}

\author{
N. \,Sanna\inst{1} \&  R. \,Contreras Ramos\inst{1}
}

  \offprints{N. Sanna \& R. Contreras Ramos}

\institute{
Dipartimento di Fisica e Astronomia, Universit\`a di Bologna\\
\email{nicoletta.sanna2@unibo.it   rodrigo.contreras@oabo.inaf.it}
}

\authorrunning{Sanna \& Contreras Ramos}

\titlerunning{BSS in UV Survey}

\abstract{We have used a combination of high-resolution \textit{Hubble Space 
Telescope} WFPC2 and wide-field ground-based observations in ultraviolet and 
optical bands to study the BSS population of the 
globular clusters NGC 6229 and M75. The combination of different filters 
allows us to optimally select specific stellar populations. In particular, 
the UV bands are ideal to study the hot objects such as HB and BSS stars. 
In both these clusters, BSS are more segregated than the normal stars. 
In NGC 6229 the BSS projected radial distribution is found to be bimodal, 
with a high central peak, a well defined minimum at intermediate radii, and 
an upturn in the outskirts, while no significant upturn in the BSS frequency 
has been observed in the outskirts of M75, suggesting that these clusters are 
in different dynamical evolutionary phases.
}
\maketitle{}

\section{Introduction}

In the optical colour-magnitude diagram (CMD) of globular clusters
(GCs) the so called blue stragglers stars (BSS) are bluer and brighter
than the main sequence objects, appearing younger and more
massive than the normal cluster stars (as also confirmed by direct
mass measurements, e.g. Shara et al. 1997). 
Two main scenarios have been proposed to explain the formation of BSS: 
mass transfer in binary systems (McCrea 1964) and stellar collisions (Bailyn 1995).
Due to mass segregation, BSS are expected to preferentially populate
the innermost region of star clusters. 
Because of stellar crowding, the
acquisition of complete samples of BSS in the core of GCs is a quite difficult
task in the optical bands. Conversely, it is easy in the UV bands (Paresce et
al. 1991).
Thanks to the advent of the \textit{Hubble Space Telescope} (HST) coupled with
wide-field imagers on ground-based telescopes, it has become possible to survey
the BSS population over the entire extension of GCs. Taking advantage of these
possibilities, the BSS radial distribution in several GCs has been
carefully analyzed by our group (Ferraro et al. 2012 and references therein).
In the majority of the clusters the radial distribution shows a bimodal behaviour, 
with a high peak in the centre, a minimum at intermediate radii and an upturn in the outer regions.
However, in some cases the distribution has been found to follow a different pattern: a flat distribution over the
entire extension of the cluster (see for example the case of $\omega$ Centauri, Ferraro et al. 2006) or an unimodal distribution without the
external rising branch (Lanzoni et al. 2007). These different distributions suggest that the clusters are in different dynamical evolutionary phases (Ferraro et al. 2012).\\
Here we present our most recent results on the BSS radial distribution of two different
clusters: NGC~6629 and M75.

\begin{table*}
\begin{center}
\caption{List of clusters included in the WFPC2 UV Survey. The flags indicate the clusters for which GALEX ($^1$) or SBC/ACS ($^2$)
images are available.}
\begin{tabular}{llllll}
\hline
NGC 104$^1$&NGC 4147$^{1,2}$&NGC 5694$^1$&NGC 6229$^{1,2}$&NGC 6402&NGC 6809$^1$\\
NGC 288$^1$&NGC 4590$^{1,2}$&NGC 5824$^1$&NGC 6254$^1$&NGC 6624$^1$&NGC 6838\\
NGC 362$^1$&NGC 4833&NGC 5904&NGC 6266$^1$&NGC 6626&NGC 6864$^1$\\
NGC 1261$^{1,2}$&NGC 5024$^{1,2}$&NGC 6093$^2$&NGC 6284$^2$&NGC 6656&NGC 6934\\
NGC 1851$^1$&NGC 5053&NGC 6121&NGC 6293&NGC 6681&NGC 6981$^1$\\
NGC 1904$^1$&NGC 5139&NGC 6171&NGC 6341$^1$&NGC 6723&NGC 7078\\
NGC 2298$^2$&NGC 5272$^1$&NGC 6205&NGC 6388$^2$&NGC 6752&NGC 7089$^1$\\
NGC 2808&NGC 5466$^1$&NGC 6218$^1$&NGC 6397$^2$&NGC 6779$^1$&\\
\hline
\end{tabular}
\end{center}
\end{table*}

\section{Strategy}
As part of the COSMIC-LAB project focused on the 
investigation of exotic objects in GCs, we obtained several images collected with the Wide Field Planetary Camera 2
(WFPC2) on board the HST.
The WFPC2 UV Survey (Prop. 11975, PI: Ferraro) focused on the study of BSS and include 47 GCs listed in Table 1. All these clusters 
have been observed in optical bands and in at least one UV WFPC2 filters ($F160BW$, $F170W$, $F218W$,
$F255W$). For some of them there are also images taken with at least one of
the UV filters of the SBC/ACS ($F140LP$, $F150LP$, $F165LP$) and with GALEX in
both the NUV and FUV filters.
Here we present the case of NGC 6229, as an example of the strategy used to reduce the data and to
select the stars. The same approach was used in the case of M75 (see Sanna et
al. 2012 and Contreras Ramos et al. 2012, for more details).
In order both to resolve the stars in the crowded central regions, and
to cover the entire extension of the cluster, we combined our high-resolution WFPC2 data with wide-field archive images. 
All the data have been reduced with the DAOPHOT/ALLFRAME packages (Stetson 1987,
1994). As shown in Figures 1 and 2 of 
Sanna et al. (2012, see also Figure 2 of Contreras Ramos et al. 2012), we have divided the entire field of view (FOV) in two samples: 
the \textit{HST sample} and the \textit{External sample}. The first one includes 
all the stars in the HST FOV within $r=90''$, while the other includes all the 
stars with $r>90''$.

\subsection{The populations selection} 
Optical images of GCs
are dominated by cool bright red giants, which blend together, preventing the 
measure of hot (faint) objects. In the UV images, instead, the brightest objects are horizontal branch (HB) 
stars and BSS and blending effects are much less common even in the very central 
region of the cluster. Accordingly, hot objects
are easily identifiable and accurately measurable (see for example Figure 1 of Contreras Ramos et al. 2012). 
For these reasons the combination 
of the UV and the optical images allows us to well identify all the stellar populations of the cluster.\\
For the \textit{HST sample} we selected the BSS population in the
($m_{255},m_{255}-m_{336}$) diagram. 
The adopted selection box is
shown in Figure~1.
These selected stars
are also shown in the left panel of Figure~2 (gray triangles). 
To select the BSS in the \textit{External sample} we 
used the same magnitude limits in
the optical plane drawn by stars selected in the UV, as shown
in grey in the right panel of the figure. 
In order to study the BSS properties we need to select also a
reference population representative of the normal cluster stars. To
this end, we consider both the HB and the red giant branch (RGB) populations. 
We decided to select HB stars by using both the UV
and the optical planes, to select the blue and the red objects.
We first selected HB stars in the UV plane by
using the selection box shown in Figure~1. Then we identified them in the ($m_{555}$, $m_{336}-m_{555}$) 
plane and we built a selection box in the
optical CMD including both these stars and the red HB
stars (see left panel of
Figure~2), allowing us to identify all the HB stars and
to obtain the magnitude limits for the selection in the outermost
regions of the cluster.  In fact,
for the \textit{External sample} 
we selected HB stars in the
($m_{555},m_{555}-r$) CMD by using the same magnitude limits
used in the optical CMD of the \textit{HST sample}. 
To select the RGB stars we used the optical
CMDs, where these objects are bright and the branch well defined. 
In order to reduce the contamination from subgiant and asymptotic giant branch stars,
we have excluded the brightest and the faintest objects. 
The boxes adopted to identify the reference
populations are shown in both panels of Figure~2.

\begin{figure}
\includegraphics[scale=0.3]{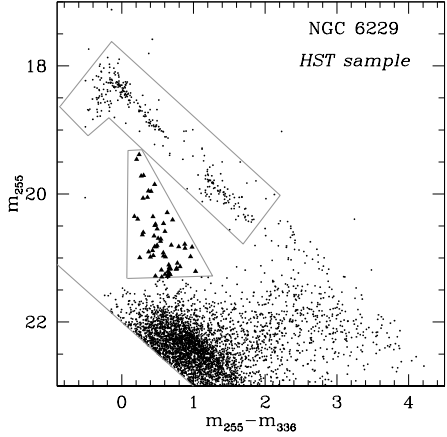}
\caption{NGC 6229: UV CMD of the \textit{HST sample}. The adopted boxes used for
  the selection of the BSS (triangles) and the HB populations
  are shown.}
\end{figure}

\begin{figure}
\includegraphics[scale=0.3]{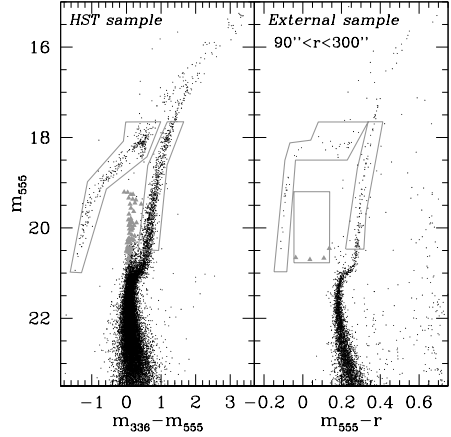}
\caption{NGC 6229: Optical CMDs of the \textit{HST} and the \textit{External}
  samples. The adopted selection boxes for BSS, RGB and HB stars are
  shown.}
\end{figure}

We computed the population number count ratios
$N_{\rm {BSS}}/N_{\rm {HB}}$, $N_{\rm {BSS}}/N_{\rm {RGB}}$ and
$N_{\rm {HB}}/N_{\rm {RGB}}$ in several concentric annuli as a function of the radial
distance from the cluster centre.
In order to ensure that our results are not affected by
severe field contamination, we carefully evaluated the expected number
of field stars in each selection box. To this end we exploited the 
FOV covered by the External catalogue, that allowed us to statistically
quantify the contamination of the field stars well beyond the tidal radius.

\section{Conclusions}

The $N_{\rm {HB}}/N_{\rm {RGB}}$ ratio both in NGC 6629 and M75 (bottom panels of Figures 3) is flat
across the entire extension of the cluster, as expected for these populations. 
In both the clusters the BSS are more segregated than the reference stars.
In particular, we have identified 64 BSS in NGC 6229 and their distribution (shown in the left panels of the figure)
is clearly bimodal, with a high peak in the centre, a minimum at
$r\sim40''$ and a rising branch in the outer region. This suggests that 
the central regions of NGC~6229 likely are already relaxed, while its outskirts 
are still not much affected by dynamical friction effects, as found also in the majority of the cluster investigated so far 
(Ferraro et al. 2012).
In the case of M75, 62 BSS have been identified. 
As shown in the right panels of the figure, these stars are more segregated in the centre, but no significant upturn of the distribution 
at large radii 
has been detected, suggesting that dynamical friction has efficiently worked in shaping the distribution of 
the BSS population in the entire cluster. This distribution is similar to that discovered by Lanzoni et al. 
(2007) in M79.
The two different BSS radial distributions suggest that M75 and NGC 6229 are in different evolutionary phases.
In particular, NGC 6229 seems to be dynamically younger than M75 considering the different efficiency of  
friction effects. The cases of $\omega$ Centauri (Ferraro et al. 2006), NGC 2419 (Dalessandro et al. 2008) and Palomar 14
(Beccari et al. 2011) support this emerging scenario. 
The BSS radial distribution of
these three objects is flat along the entire extension of the cluster, suggesting that these clusters are not
relaxed yet. This scenario suggests that we can roughly divide the clusters in
at least three main groups: young (not relaxed yet), 
intermediate (with a bimodal BSS distribution) and dynamically old (with an unimodal BSS distribution) GCs.
Further BSS surveys covering the full radial extent of GCs will be a powerful tool to confirm this emerging scenario
and will help to extend our knowledge of BSS distribution and its connection to the cluster dynamical history.

\begin{figure}
\includegraphics[height=0.4\textheight,width=0.5\textwidth]{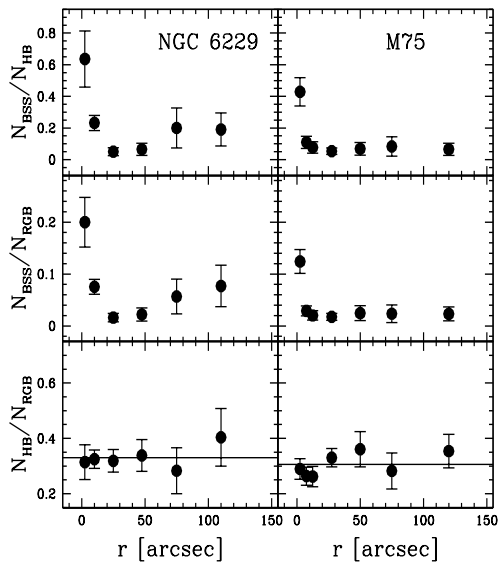}
\caption{Distribution of the population ratios as a function of the
  radial distance from the cluster centre of NGC 6229 (left) and M75 (right).}
\end{figure}

\begin{acknowledgements}
This research is part of the project COSMIC-LAB funded by the
European Research Council (under contract ERC-2010-AdG-267675).
\end{acknowledgements}

\bibliographystyle{aa}

\end{document}